\documentstyle[aps,multicol,epsfig]{revtex}

\def\<{\langle}
\def\>{\rangle}

\newcommand{\be}{\begin{equation}}
\newcommand{\ee}{\end{equation}}
\newcommand{\bea}{\begin{eqnarray}}
\newcommand{\eea}{\end{eqnarray}}

\begin{document}


\title{An example of the difference between quantum and classical random
walks}
\author{Andrew M. Childs,$^1$ Edward Farhi,$^1$ and Sam Gutmann$^2$}
\address{$^1$ Center for Theoretical Physics, Massachusetts Institute of
              Technology, Cambridge, MA 02139 \\
         $^2$ Department of Mathematics, Northeastern University, Boston, 
              MA 02115}
\date{5 March 2001}

\maketitle


\begin{abstract}
In this note, we discuss a general definition of quantum random walks on
graphs and illustrate with a simple graph the possibility of very different
behavior between a classical random walk and its quantum analogue.  In this
graph, propagation between a particular pair of nodes is exponentially
faster in the quantum case.
\hfill [MIT-CTP \#3093]
\end{abstract}

\begin{multicols}{2}

\noindent {\bf Introduction.}
Many classical algorithms are based on random walks, so it is natural to ask
whether quantum random walks might be useful for quantum computation.  A
framework for using quantum random walks to solve decision problems was
investigated in~\cite{FG98}.  There also, an exponential separation was
found between the classical and quantum times to propagate through a certain
tree.

In this note, we describe a general definition of continuous-time random
walks on graphs and give a simpler example of a graph for which the quantum
time to propagate between a particular pair of nodes is exponentially
shorter than the analogous classical propagation time.  We also discuss
advantages of the continuous time formulation over discrete versions.

\medskip\noindent {\bf Random walks.}
A continuous time classical random walk on a graph is a Markov process.  A
graph is a set of $v$ vertices $\{1, 2, \ldots, v\}$ and a set of edges that
specifies which pairs of vertices are connected in the graph.  A step in a
classical random walk on a graph only occurs between two vertices connected
by an edge.  Let $\gamma$ denote the jumping rate.  Starting at any vertex,
the probability of jumping to any connected vertex in a time $\epsilon$ is
$\gamma \epsilon$ (in the limit $\epsilon \to 0$).  This random walk can be
described by the $v \times v$ infinitesimal generator matrix $M$ defined by
\be
  M_{ab} = \left\{ 
    \begin{array}{ll}
      -\gamma & \textrm{$a \ne b$, $a$ and $b$ connected by an edge} \\
      0       & \textrm{$a \ne b$, $a$ and $b$ not connected} \\
      k\gamma & \textrm{$a=b$, $k$ is the valence of vertex $a$.}
    \end{array} \right.
\ee
If $p_a(t)$ denotes the probability of being at vertex $a$ at time $t$, then
\be
  {{\mathrm d} p_a(t) \over {\mathrm d}t} = - \sum_b M_{ab} \, p_b(t)
\,.
\label{eq:diffeq}
\ee

Consider quantum evolution in a $v$-dimensional Hilbert space according to a
Hamiltonian $H$.  In a basis $|1\>, |2\>, \ldots, |v\>$, the Schr\"odinger
equation for $|\psi(t)\>$ can be written
\be
  i {{\mathrm d} \over {\mathrm d}t} \<a|\psi(t)\> 
    = \sum_b \<a|H|b\> \<b|\psi(t)\>
\,.
\label{eq:schrodinger}
\ee
Note the similarity between (\ref{eq:diffeq}) and (\ref{eq:schrodinger}).
Whereas (\ref{eq:diffeq}) conserves probability in the sense that
\be
  \sum_a p_a(t) = 1
\,,
\ee
the Schr\"odinger equation preserves probability as the sum of the
amplitudes squared:
\be
  \sum_a |\<a|\psi(t)\>|^2 = 1
\,.
\ee
In some sense, {\em any} evolution in a finite-dimensional Hilbert space can
be thought of as a ``quantum random walk.''  However, the analogy is
clearest when $H$ has an obvious local structure.

A quantum random walk on a graph is naturally defined in a Hilbert space
spanned by basis elements corresponding to the vertices.  To respect the
structure of the graph, we require that for $a \ne b$,
\be
  \<a|H|b\> \ne 0~\textrm{iff $a$ and $b$ are connected by an edge.}
\ee
This is a very weak requirement, so we can impose more structure on $H$.  A
natural quantum analogue to the classical random walk described above is
given by the quantum Hamiltonian with matrix elements~\cite{FG98}
\be
  \<a|H|b\> = M_{ab}
\,.
\label{eq:ham}
\ee
Note that on a one-dimensional lattice, this results in the Hamiltonian
defined by
\be
  H|j\> = -{1 \over \Delta^2}(|j-1\>-2|j\>+|j+1\>)
\,,
\ee
which is just a discrete approximation to the operator $-{\mathrm
d}^2/{\mathrm d}x^2$ (where $\Delta=\gamma^{-1/2}$ is the lattice spacing).

The difference between the quantum and classical evolution comes from the
$i$ which appears in (\ref{eq:schrodinger}) but not in (\ref{eq:diffeq}).
This can result in radically different behavior, as seen in~\cite{FG98}.  A
simpler example is given next.

\begin{figure}
  \begin{center}
  \psfig{figure=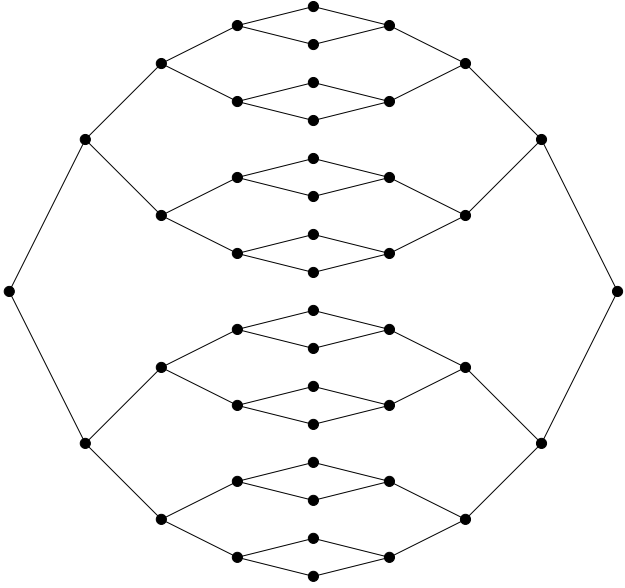,width=3.25in}
  \end{center}
  \caption{The graph $G_4$.}
  \label{fig:graph}
\end{figure}

\medskip \noindent {\bf An example.}
Here we define a sequence of graphs $G_n$.  The number of vertices in $G_n$
is $2^{n+1}+2^n-2$.  In Figure~\ref{fig:graph} we show $G_4$.  In general,
$G_n$ consists of two balanced binary trees of depth $n$ with the $2^n$
$n$th-level vertices of the two trees pairwise identified.

For both the classical and quantum random walks, we start at the root of one
tree and want the probability as a function of time of being at the other
root.  In other words, we are interested in how long it takes to propagate
from the leftmost vertex to the rightmost vertex as a function of $n$.

Consider the classical case first.  The vertices of $G_n$ can be grouped in
columns indexed by $j \in \{0, 1, \ldots, 2n\}$.  Column $0$ contains the
root of the left tree, column $1$ contains the two vertices connected to
that root, etc.  Note that column $n$ contains the $2^n$ vertices in the
middle of the graph and column $2n$ is the root at the right.

To analyze the classical walk from the left root to the right root, we need
only keep track of the probabilities of being in the columns.  In the left
tree, for $0<j<n$, the probability of stepping from column $j$ to column
$j+1$ is twice as great as the probability of stepping from column $j$ to
column $j-1$.  However, in the right tree, for $n<j<2n$, the probability of
stepping from column $j$ to column $j+1$ is half as great as the probability
of stepping from column $j$ to column $j-1$.  This means that if you start
at the left root, you quickly move to the middle of the graph, but then it
takes a time exponential in $n$ to reach your destination.  More precisely,
starting in column $0$, the probability of being in column $2n$ after any
number of steps is less than $2^{-n}$.  This implies that the probability of
reaching column $2n$ in a time that is polynomial in $n$ must be
exponentially small as a function of $n$.

We now analyze the quantum walk on $G_n$ starting in the state corresponding
to the left root and evolving with the Hamiltonian given by (\ref{eq:ham}).
With this initial state, the symmetries of $H$ keep the evolution in a
$(2n+1)$-dimensional subspace of the $(2^{n+1}+2^n-2)$-dimensional Hilbert
space.  This subspace is spanned by states $|{\mathrm col}~j\>$ (where $0
\le j \le 2n$), the uniform superposition over all vertices in column $j$,
that is,
\be
  |{\mathrm col}~j\> 
    = {1 \over \sqrt{N_j}} \sum_{a \in {\mathrm column}~j} |a\>
\,,
\ee
where
\be
  N_j = \left\{ \begin{array}{ll}
    2^j      & 0 \le j \le n \\
    2^{2n-j} & n \le j \le 2n \,.
    \end{array} \right.
\ee

\begin{figure}
  \begin{center}
  \psfig{figure=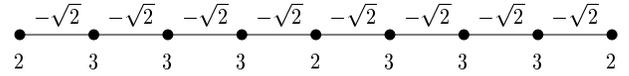,width=3.25in}
  \end{center}
  \caption{The reduction of $G_4$ to a quantum random walk on a line.
  Vertices correspond to columns and are labeled with the diagonal matrix
  elements of $H/\gamma$, whereas edges are labeled with its matrix elements
  between adjacent columns.}
  \label{fig:line}
\end{figure}

In this basis, the non-zero matrix elements of $H$ are
\bea
  \<{\mathrm col}~j|H|{\mathrm col}~j\pm1\> &=& -\sqrt{2}\gamma \\
  \<{\mathrm col}~j|H|{\mathrm col}~j\> &=& \left\{ \begin{array}{ll}
    2\gamma & j=0,n,2n \\
    3\gamma & {\mathrm otherwise,} \\
    \end{array} \right.
\eea
which is depicted in Figure~\ref{fig:line} (for $n=4$) as a quantum random
walk on a line with $2n+1$ vertices.

\begin{figure}
  \begin{center}
  \psfig{figure=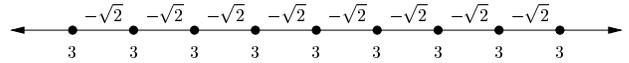,width=3.25in}
  \end{center}
  \caption{Quantum random walk on an infinite, translationally invariant
  line.}
  \label{fig:infline}
\end{figure}

Starting at the leftmost vertex of Figure~\ref{fig:line}, there is an
appreciable probability of being at the rightmost vertex after a time
proportional to $n$.  To see this, first consider quantum propagation on an
infinite, translationally invariant line of vertices as depicted in
Figure~\ref{fig:infline}.  Here it is straightforward to compute the
amplitude to go from vertex $l$ to vertex $m$ in a time $t$ (for example,
see~\cite{FG92}):
\be
  \<m|e^{-iHt}|l\> = e^{-i 3 \gamma t} i^{m-l} J_{m-l}(2\sqrt2 \gamma t)
\,,
\ee
where $J_{m-l}$ is a Bessel function of order $m-l$.  This corresponds to
propagation with speed $2\sqrt2 \gamma$.  More precisely, for any
$\epsilon>0$ and $|m-l| \gg 1$, for $t < \left({1 \over 2 \sqrt2
\gamma}-\epsilon\right)|m-l|$, the amplitude is exponentially small in
$|m-l|$, whereas there are values of $t$ between $({1 \over 2 \sqrt2
\gamma})|m-l|$ and $({1 \over 2 \sqrt2 \gamma}+\epsilon)|m-l|$ at which the
amplitude is of order $|m-l|^{-1/2}$.

\begin{figure}
  \begin{center}
  \psfig{figure=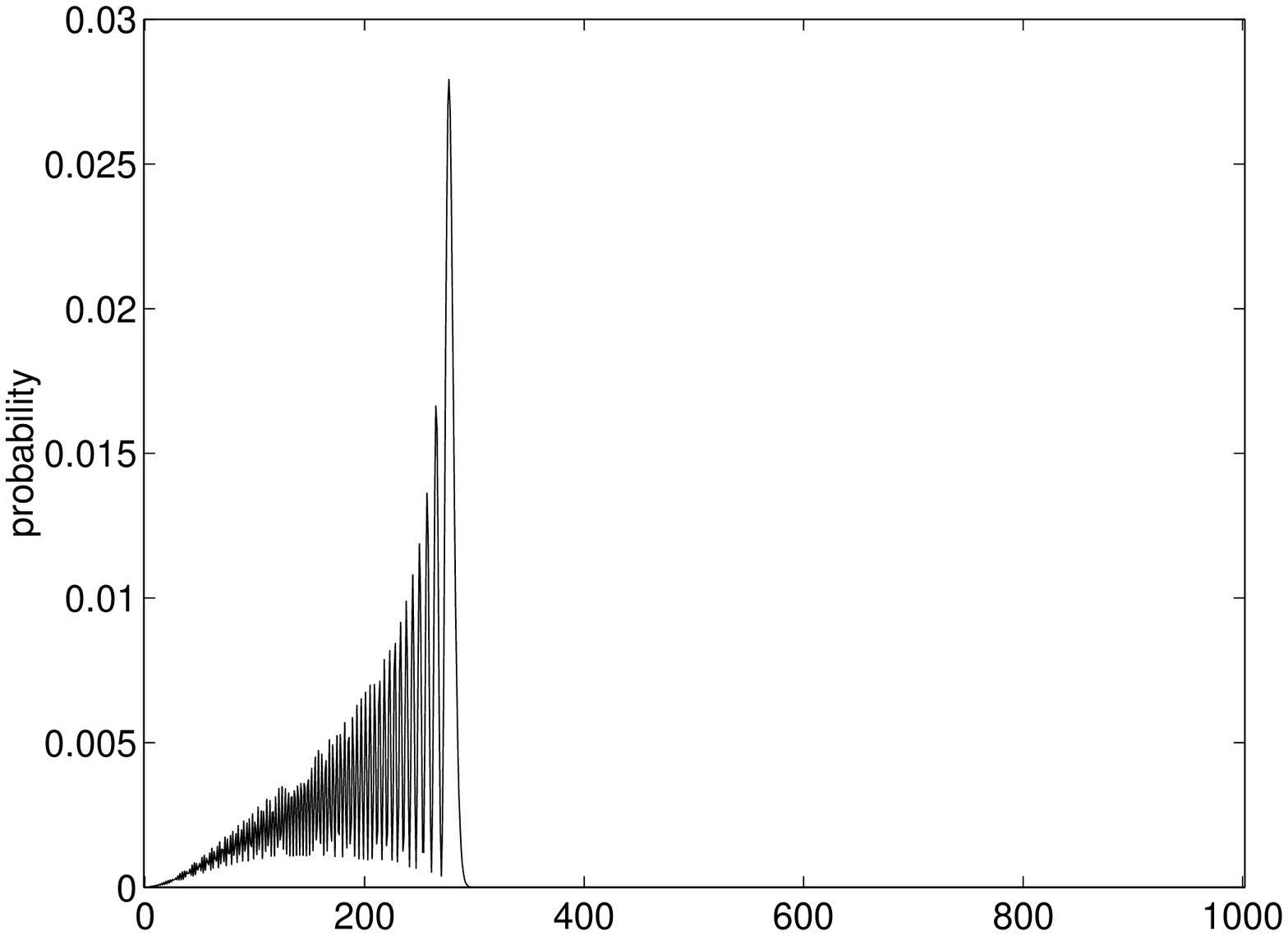,width=3.25in}
  \psfig{figure=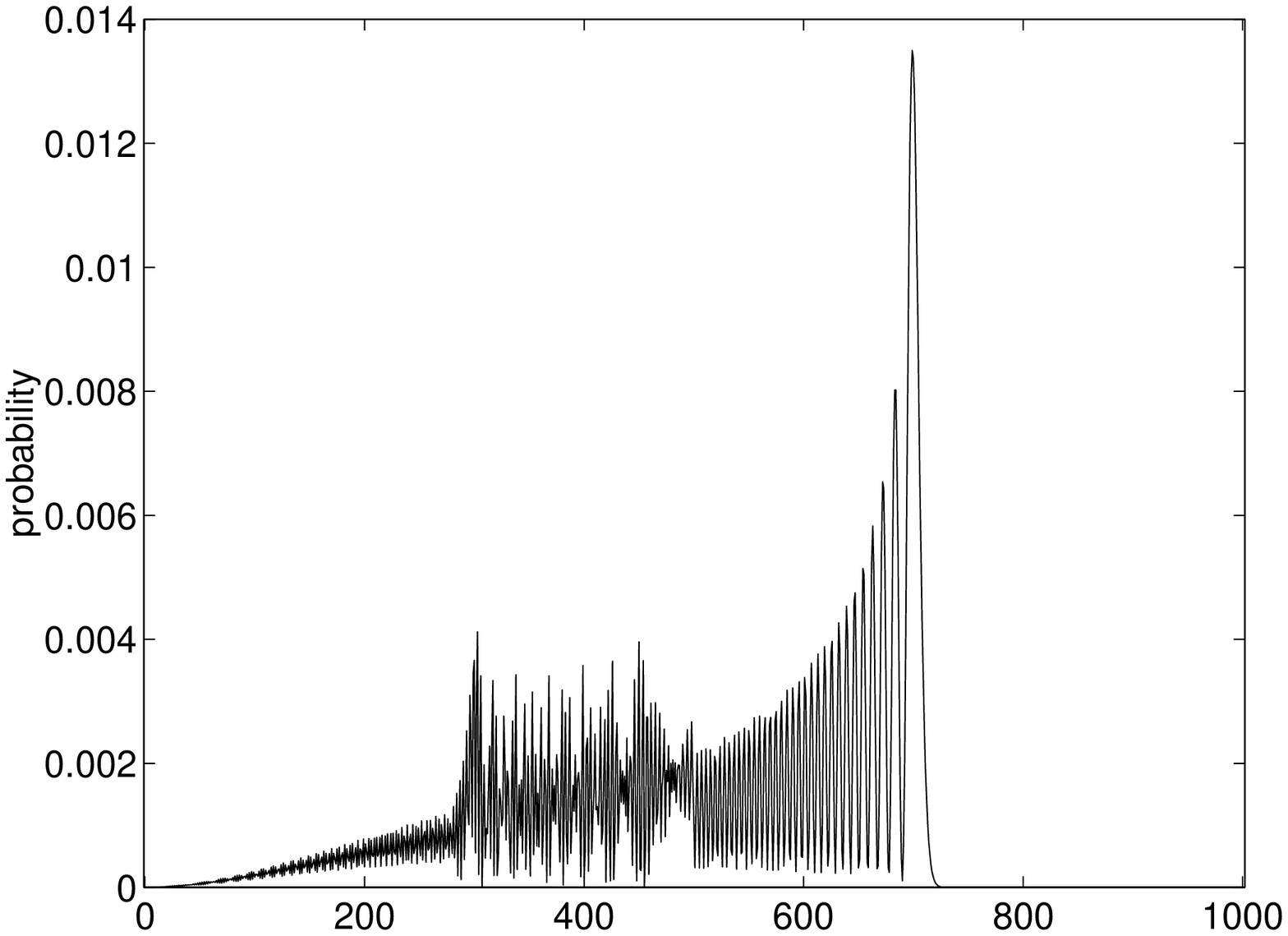,width=3.25in}
  \psfig{figure=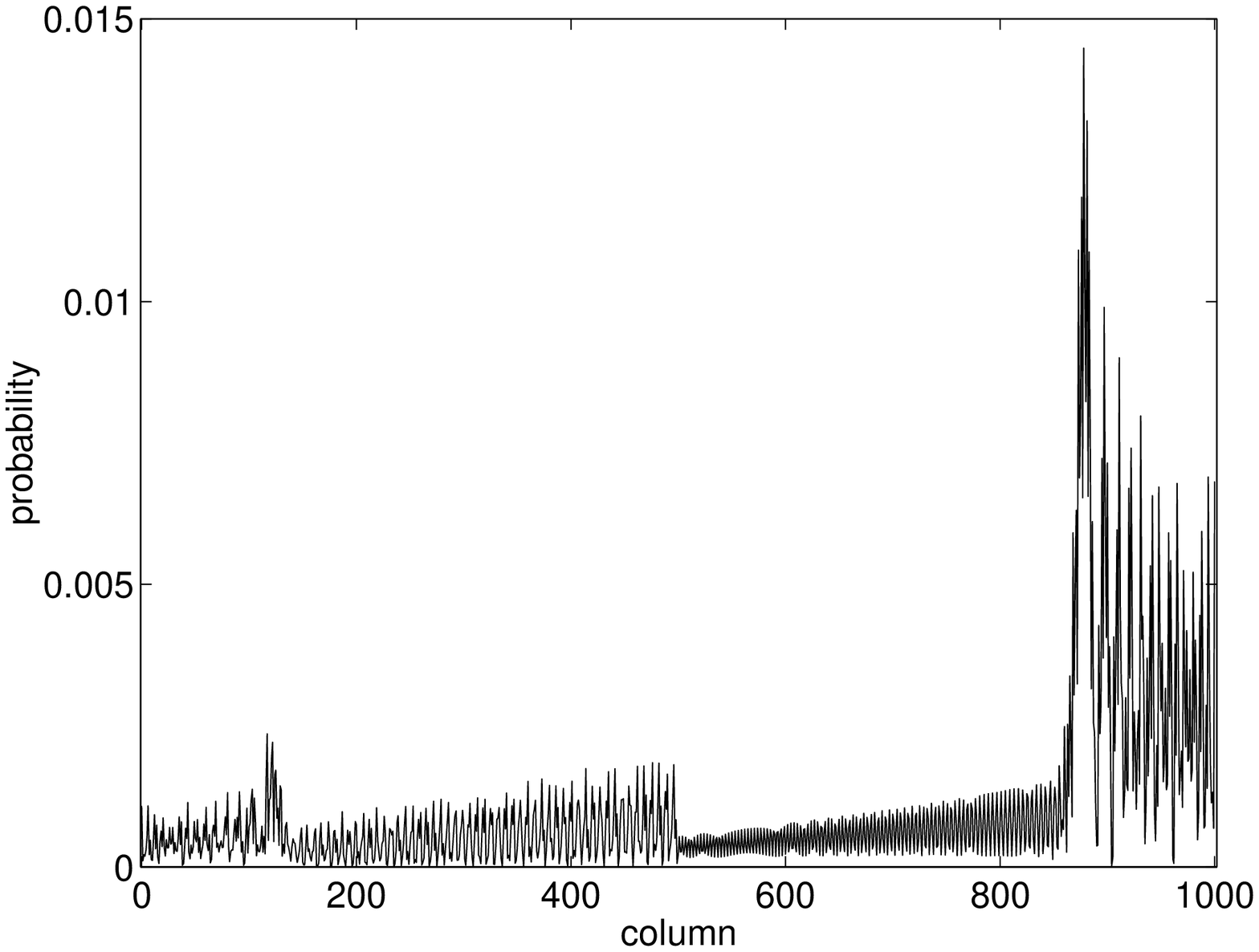,width=3.25in}
  \end{center}
  \caption{Propagation in $G_{500}$ starting at the left root.  From top to
  bottom, the times are $t=100$, $250$, and $400$.}
  \label{fig:propagation}
\end{figure}

In the limit of large $n$, the reduced version of $G_n$ is nearly identical
to the infinite, translationally invariant line, so it is plausible that
propagation on $G_n$ will also occur with speed $2\sqrt2\gamma$.  To verify
this, we numerically compute the probability $|\<{\mathrm
col}~j|\psi(t)\>|^2$ of being in column $j$ at various times $t$, where
$|\psi(0)\>=|{\mathrm col}~0\>$ and we choose $\gamma=1$.  This is shown in
Figure~\ref{fig:propagation} with $n=500$ for $t=100$, $250$, and $400$.
These plots clearly show a wave packet which propagates with speed
$2\sqrt2$, with the amplitude near the wavefront decreasing like $t^{-1/2}$.
In the first plot, at $t=100$, the leading edge of the distribution is at
column $200\sqrt2 \approx 283$.  The packet has not yet encountered the
small defect at the center, so it has a relatively simple shape.  At
$t=250$, the wavefront has passed the center, and a small reflection can be
seen propagating backward.  However, the leading edge is relatively
undisturbed, having propagated to column $500\sqrt2 \approx 707$.  The
wavefront continues to propagate with speed $2\sqrt2$ until it reaches the
right root, where the packet is reflected.  The last plot, at $t=400$, shows
the distribution shortly after this first reflection.  Even after the
reflection, there is still an appreciable probability of being at the right
root.

\medskip \noindent {\bf The limiting distribution.}
In this section, we consider the distribution over the vertices after a long
time.  We emphasize that although the mixing times (the characteristic times
to reach the limiting distribution) may be similar in the classical and
quantum cases~\cite{AAKV00}, this is in no way indicative of similar
dynamics, as the limiting distributions may be radically different.

In the classical case, the limiting distribution is defined as
\be
  \pi_b = \lim_{T \to \infty} p_b(T)
\,,
\ee
which is independent of the starting state.  It is easy to see that the
limiting distribution on $G_n$ is uniform over the vertices: this
distribution is the unique eigenvector of $M$ with eigenvalue $0$, so it is
the only component that remains after a long time.  Thus $\pi_b =
(2^{n+1}+2^n-2)^{-1}$ for each vertex $b$, which is exponentially small.

In the quantum case, unitarity prevents the walk from reaching a steady
state.  However, a sensible definition of the limiting distribution, which
depends on the initial state $|a\>$, is given by~\cite{AAKV00}
\be
  \chi_b = \lim_{T \to \infty} {1 \over T} \int_0^T |\<b|e^{-iHt}|a\>|^2 
           \, {\mathrm d}t
\,.
\ee
This is the distribution resulting from a measurement done after a time
chosen uniformly in $[0,T]$, in the limit of large $T$.  By expanding over
the energy eigenstates $|E_r\>$, we find
\bea
  \chi_b &=& \sum_{r,s} \<b|E_r\>\<E_r|a\>\<a|E_s\>\<E_s|b\> \nonumber\\
         &&  \quad\times \lim_{T \to \infty} {1 \over T} 
             \int_0^T e^{-i(E_r-E_s)t} \, {\mathrm d}t \\
         &=& \sum_r |\<a|E_r\>|^2 \, |\<b|E_r\>|^2
\eea

\noindent
(note that we have assumed $E_r \ne E_s$ for $r \ne s$, which is true for
$G_n$).  In particular, consider the case where $|a\>=|{\mathrm col}~0\>$
corresponds to the left root and $|b\>=|{\mathrm col}~2n\>$ corresponds to
the right root.  In this case, we may work in the reduced Hilbert space
spanned by the columns, so the number of energy eigenstates is $2n+1$.  By
symmetry, $|\<{\mathrm col}~0|E_r\>| = |\<{\mathrm col}~2n|E_r\>|$.  The
Cauchy-Schwartz inequality gives
\be
  \sum_r |\<{\mathrm col}~0|E_r\>|^4 \, \sum_s 1 
  \ge \left(\sum_r |\<{\mathrm col}~0|E_r\>|^2\right)^2 = 1
\,,
\ee
which implies
\be
  \sum_r |\<{\mathrm col}~0|E_r\>|^4 \ge {1 \over 2n+1}
\,.
\ee
Thus in the limiting distribution, the probability of being at the right
root, starting at the left root, is
\be
  \chi_{{\mathrm col}~2n} \ge {1 \over 2n+1}
\,,
\ee
which is much larger than in the classical case.

\medskip \noindent {\bf Discussion.}
The model of quantum random walks used in this note applies automatically to
any graph.  In particular, the Hamiltonian is determined by the local
structure of the graph and its definition does not require knowledge of any
global properties.  It is easy to imagine situations where the local
structure of a graph is readily accessible, but determining some global
property is difficult.  For example, a computational problem may involve
searching a graph for a node with a certain property whose presence or
absence from the graph corresponds to the solution of an NP-complete
problem~\cite{FG98}.

The Hamiltonian-based approach to quantum random walks can be contrasted
with discrete time models (for example, see~\cite{AAKV00,ADZ93,NV00})
involving the extra state space of a ``quantum coin.''  This extra label
seems to be necessary in discrete time formulations of quantum random walks
(and is provably necessary in the one dimensional case~\cite{M96}).
However, for general graphs of mixed valence, it is not obvious how to
define the discrete time unitary evolution operator without knowledge of
global properties of the graph.

\medskip \noindent {\bf Acknowledgements.}
This work was supported in part by the Department of Energy under
cooperative agreement DE-FC02-94ER40818.  AMC is supported by the Fannie and
John Hertz Foundation.


\end{multicols}
\end{document}